\documentstyle[11pt]{article}

\title{\bf On the triple pomeron contribution in the hard pomeron theory}
\author { M.A.Braun\thanks{Visiting professor IBERDROLA; on
leave of absence from the  Department of High Energy Physics,
University of St. Petersburg, 198904 St. Petersburg, Russia.}\\ Departamento
de F\'{\i}sica de Part\'{\i}culas,\\ Universidade de Santiago de
Compostela,\\ 15706-Santiago de Compostela, Spain}
\date{April 1996}
\def\beq{\begin{equation}}
\def\eeq{\end{equation}}
\def\noi{\noindent}

\begin{document}
\maketitle
\medskip
\noi{\bf Abstract.}
Within the hard pomeron approach it is shown that the triple pomeron
interaction gives no contribution to the cross-section for a  projectile
with large enough virtuality $g^{2}\ln Q^{2}\sim 1$. For small virtualities
$g^{2}\ln Q^{2}<<1$ its contribution is essential.
\vspace{7.0cm}

\noi{\Large\bf hep-ph/9604350}\\
\noi{\Large\bf US-FT/17-96}
\newpage
\section{Introduction}
Naive estimates seem to show that the triple pomeron diagram contributes
significantly to the scattering cross-section at high energies. 
Take the
diagram of Fig. 1$a$ as an example. Let the total c.m. energy squared be
$s=\exp Y$,  those for the upper and lower pomerons $s_{1}=\exp y_{1}$ and
$s_{2}=\exp y_{2}$, respectively, $y_{1}+y_{2}=Y$,  and $\Delta$ the pomeron
intercept. Then the contribution of the diagram seems to be
\beq
\gamma_{1}\gamma_{2}\gamma_{3P}\int_{0}^{Y}dy_{1}\exp (\Delta y_{1})
\exp (2\Delta (Y-y_{1}))=\gamma_{1}\gamma_{3}\gamma_{3P}\Delta^{-1}
\exp(2\Delta Y)(1-\exp(-\Delta Y)
\eeq 
where $\gamma_{1}$ and $\gamma_{2}$ are the couplings of the colliding
particle (the same for the projectile and target, for simplicity) to one
and two pomerons and $\gamma_{3P}$ is the triple pomeron coupling.
This contribution is of the same order as the one from the "pure" 
two pomeron exchange, corresponding to the diagram of Fig. 1$b$
\beq
\gamma_{2}^{2}\exp (2\Delta Y)
\eeq
even in the small coupling limit, when one should take into account that
$\gamma_{3P}\sim\gamma_{2}$.

However inspecting the result (1) one notices that the whole contribution
to the righthand side at large $Y$ comes from the region of integration on
the left $y_{1}<<Y$, where the existence of the upper pomeron becomes
questionable. So the conclusion that at large energies the triple pomeron
gives a sizable contribution is not at all obvious. In fact, in our earlier
paper [1] we assumed that it does not contribute, which lead us to the
cross-section essentially of the eikonal form.

In this note we try to study the triple pomeron contribution in
the framework of the BFKL hard pomeron [2]  in some detail. We shall study
only the simplest diagram of Fig. 1$a$ and even for this diagram we shall
not be able to obtain results  in a form, amenable for numerical
calculations. However, what we shall find is sufficient to draw definite
conclusions as to the relative importance of the triple pomeron
contribution. Since the derivation is very technical, we present here the
main result together with its simple explanation.

As it turns out, the relative importance of the triple pomeron contribution
depends on the magnitude of the colliding particle virtuality (or mass)
$Q^{2}$. If it is high enough (e.g. for the structure function) then the
contribution of the triple pomeron is neglegible. However, with $Q^{2}$
diminishing, the triple pomeron contribution becomes more important and at
low $Q^{2}$ it results at least of the same magnitude as the pure two
pomeron exchange (Fig. 1$b$).

To understand this result note that the hard pomeron theory with a small
(fixed) coupling constant $g$ possesses an intrinsic large rapidity (or
energy) scale
\beq Y_{0}=1/g^{2} \eeq
This should be contrasted to the "external" rapidity $Y$ and colliding
particle virtuality $Q^{2}=\ln\zeta$. Evidently the pomeron theory applies
only when
\beq
1<<Y_{0}<<Y,\ \ \zeta <<Y
\eeq
However the relation between $\zeta$ and $Y_{0}$ may be arbitrary.

 Now take
the generic diagram of Fig. 2 corresponding to  both of the diagrams in Fig.
1, with the upper blob $B$ taken at the energy $s_{1}=\exp y_{1}$. Its
contribution is given by the integral similar to (1)
\beq
\gamma_{2}\int_{0}^{Y}dy_{1}B(y_{1},\zeta)
\exp (2\Delta (Y-y_{1}))
\eeq 
The integration in (5) is evidently limited to  values of $y_{1}$
smaller or of the order of $\Delta^{-1}\sim Y_{0}$. Then we may distinguish
between two possibilities.

If $Q^{2}$ is large, so that $\zeta$ is of the order (or greater) than
$Y_{0}$
\beq \zeta\sim Y_{0} \eeq
then  blob $B$ is, in fact,  not in the low-$x$ but in a pure
perturbative regime (recall that the coupling constant is supposed to be
fixed and small). Then, to the accuracy adopted in the BFKL theory, we have
to retain only the lowest order contribution to it. This leads to the
pure double pomeron exchange. For the photon projectile it means that the two
pomerons should be directly coupled to the  the $q\bar q$ loop. Thus in the
region (6) the triple pomeron does not contribute: it simply does not exist.

On the other hand for small $Q^{2}$, i.e. when
\beq
\zeta<<Y_{0}
\eeq
blob $B$ turns out to be in the low-$x$ regime and we can take a pomeron 
for its leading contribution $B(y,\zeta)=\gamma_{1}\gamma_{3P}P(g^{2}y)$.
The integral (5) then gives
\beq
\gamma_{1}\gamma_{2}\gamma_{3P}g^{-2}\exp(2\Delta Y)
\int_{0}^{Y/g^{2}}dzP(z)\exp(-2z)
\eeq 
This contribution now does not reduce to the pure two-pomeron exchange and
corresponds to the triple p[omeron of Fig. 1$a$. Note that in (8) finite
values of $z$ contribute, so that the upper pomeron of Fig. 1$a$ enters
here not in its asymptotic regime (as in (1)) but at finite values of
$g^{2}\ln s$ and so has to be taken in its full complexity.

As a result, we find that the hard pomeron theory allows for a clear
distinction between colliding "particles", with $g^{2}\ln Q^{2}<<1$, and
virtual probes, with $g^{2}\ln Q^{2}\sim 1$. For the former the triple
pomeron dominates the cross-section. For the latter it does not contribute.

Turning to the literature, we recall that our results for summing
multipomeron exchanges in [1] referred to the photon structure function at
high $Q^{2}$. So the omission of the triple (and multiple) pomeron
interactions, which was made there, results justified. On the other hand,
in a series of papers of A.Mueller {\it et al.} [3-5] the calculation of
the double (and multiple) pomeron exchange seems to be made under the
assumption of a fixed mass $M$ of colliding "onia" and the coupling $g$
going to zero. Then one inevitably arrives at the region (7) where the triple
pomeron seems to give the bulk of the contribution. This may be a part of the
explanation why the resulting amplitudes do not eikonalize in A.Mueller's
approach.

We have to note, however, that to move closer to the realistic QCD, one has
to take the onia mass $M$ large enough to make the effective coupling
constant small. This latter condition, translated into the fixed coupling
constant language, means that $g^{2}\ln M^{2}\sim 1$, so that the region (6)
should rather be considered, where no triple pomeron contribution is to be
expected.

In the rest of this note we present the calculation of the triple pomeron
contribution (Fig. 1$a$). As mentioned we shall not be able to find an
expression for it which admits, say, numerical estimates. However we shall
clearly see how and why this contribution goes to zero when $Q^{2}$ rises.
Also a part (dominant at high $Q^{2}$) will be found in a form which can be
studied numerically. Our calculations will be heavily based on the results
obtained in our previous paper [6], where the triple pomeron vertex was
studied in the asymptotic regime.

\section{The triple pomeron contribution. The target part}
The cross-section corresponding to Fig. 1$a$. in the hard pomeron approach
of [2] is explicitly given by
\[
\sigma=-\frac{g^{10}N^{3}}{32(2\pi)^{9}}\int d^{2}l\int
(ds_{1}/s_{1})
\int \prod_{i=1}^{3}(d^{2}r_{i}d^{2}r'_{i}
(2\pi )^{-2})K_{l}(r_{1},r_{2},r_{3})\]
\beq\exp (-il(r_{2}+r_{3})/2)
G_{0}(s_{1},r_{1},r'_{1})G_{l}(s_{2},r_{2},r'_{2})
G_{-l}(s_{2},-r_{3},r'_{3})\rho_{1}(r'_{1})\rho_{2}(r'_{2},r'_{3})
\eeq
Here $N$ is the number of flavours; $s_{2}=s/s_{1}$ where $s$ is the overall
c.m. energy squared; $r'_{i}$ and $r_{i}$, $i=1,2,3$, are the initial and
final transverse dimensions of the upper ($i=1$) and lower ($i=2,3$)
pomerons as they propagate from the sources (colliding particles) to the
triple pomeron junction. The sources colour densities are $\rho_{1}$ for the
projectile and $\rho_{2}$ for the target. The BFKL Green function
$G_{l}(s,r,r')$ describes propagation of the pomeron with the total momentum
$l$ and energy $s$. $K_{l}(r_{1},r_{2},r_{3})$ is the triple pomeron
interaction vertex. It originates from the Bartels' vertex for the
transition of 2 reggeized gluons into 4 [7]. Its explicit form  in the
coordinate space was found in [6]. Taking into account that the Green
functions $G_{l}$ vanish if $r=0$ or $r'=0$, it can be reduced to
\beq
K_{l}(r_{1},r_{2},r_{3})=
-\frac{2}{(2\pi)^{2}}\frac{(r_{2}r_{3})}{r_{2}^{2}r_{3}^{2}}\nabla_{1}^{4}
\delta^{2}(r_{1}+r_{2}+r_{3})
\eeq
The numerical coefficient in (9) combines factors $g^{2}N/2$ and
$g^{4}N/2$ separated from the colour densities $\rho_{1}$ and
$\rho_{2}$, respectively, a factor $g^{4}N$ from the triple pomeron
vertex, a factor $1/2$ for the identity of two lower pomerons and some
numerical factors which relate the cross-section to the amplitude
corresponding to the diagram of Fig. 1$a$.

For our purpose the explicit form of the target density
$\rho_{2}(r'_{2},r'_{3})$ will be inessencial, except that it should be
well behaved  and restrict $r'_{2}$ and $r'_{3}$ to some average value
$r_{20}$ of the order of the target transverse dimension. To simplify, we
then take
$\rho_{2}(r'_{2},r'_{3})=
\delta^{2}(r'_{2}-r'_{3})\delta(r'_{2}-r_{20})/(2\pi
r_{20})$, which means that we simply substitute in the two Green functions
for the lower pomerons $r'_{2}=r'_{2}=r_{20}$ where the latter vector has
length $r_{20}$ and an arbitrary direction.
After this substitution, and with the explicit form of $K_{l}$ given by
(10), the cross-section (9) transforms into
\[
\sigma=-\frac{g^{10}N^{3}}{16(2\pi)^{13}}\int d^{2}l\int
(ds_{1}/s_{1})\int d^{2}r_{1}d^{2}q\rho_{1}(r_{1})\]\beq
\chi_{1}(s_{1},q+l/2,r_{1})
(\nabla_{q}\chi_{2}(s_{2},l,q))^{2}
\eeq
where
\beq
\chi_{1}(s,q,r_{1})=\int d^{2}r\nabla^{4}G_{0}(s,r,r_{1})\exp iqr
\eeq
and
\beq
\chi_{2}(s,l,q)=\int d^{2}r r^{-2}G_{l}(s,r,r_{20})\exp iqr
\eeq

As mentioned in the Introduction, the triple pomeron interaction which
enters (11) has been studied in [6] in the asymptotic region
$s_{1}\sim s_{2}\rightarrow\infty$. In our case, we rather have to
integrate over all $s_{1}$, so that the upper pomeron will stay far from
its asymptotic regime and we shall have to take the exact form for its
Green function. However the target part $\chi_{2}$ enters (11) in its
asymptotic region, so that we may borrow its form from [6], to which paper we
refer the reader for more detail in this respect.

The leading contribution
to the BFKL Green function at $l\neq 0$, which appears in (13), has the form
[8] \beq
G_{l}(s,r,r')=(1/4\pi^{2})\int\frac{d\nu \nu^{2}}{(\nu^{2}+1/4)^{2}}
s^{\omega(\nu)}E_{l}^{\nu}(r)E_{l}^{-\nu}(r')
\eeq
where
\beq
\omega(\nu)=(g^{2}N/2\pi^{2})(\psi(1)-{\rm Re}\psi(1/2+i\nu))
\eeq
is the pomeron intercept and
\beq
E_{l}^{\nu}(r)=\int d^{2}R\exp(ilR)(\frac{r}{|R+r/2||R-r/2|})^{1+2i\nu}
\eeq
 The Green function (14), transformed into momentum
space, contains terms proportional to $\delta^{2}(l/2\pm q)$, which
should be absent in the physical solution (this circumstance was first
noted by A.H.Mueller and W.-K.Tang [9]). For that, (14) goes to zero
at $r=0$. Terms proportional to $\delta^{2}(l/2+q)$ are not dangerous
to us: they are killed by the vertex $K_{l}$. 
 To remove the dangerous singularity at $q=l/2$
and simultaneously preserve a good behaviour at $r=0$, as in [6], we  make
a subtraction in $E$, changing it to
\beq
\tilde{E}^{\nu}_{l}(r)=
\int d^{2}R\exp(ilR)((\frac{r}{|R+r/2||R-r/2|})^{1+2i\nu}-
|R+r/2|^{-1-2i\nu}+|R-r/2|^{-1-2i\nu})
\eeq

Integration over $r$ leads to the integral
\beq
J_{2}(l,q)=\int (d^{2}r/(2\pi)^{2})(1/r^{2})\tilde{E}^{\nu}_{l}(r)
\eeq
This integral is convergent at any values of $\nu$, the point $\nu=0$
included, when the convergence at large values of $r$ and $R$ is
provided by the exponential factors. So in the limit
$s\rightarrow\infty$, when small values of $\nu$ dominate, we can put
$\nu=0$ in $J_{2}$:
\beq
J_{2}(l,q)=\int(d^{2}Rd^{2}r/(2\pi)^{2})\frac{\exp(ilR+iqr)}
{r|R+r/2||R-r/2|}+(1/l)\ln\frac{|l/2-q|}{|l/2+q|}
\eeq
The second term comes from the subtraction terms in (17). Passing to the
Fourier transform of the function $1/r$ one can represent (19) as an
integral in the momentum space
\beq
J_{2}(l,q)=(1/l)\int \frac{d^{2}p (l+|l/2+p|-|l/2-p|)
}{(2\pi)|l/2+p||l/2-p||q+p|}
\eeq

In the same manner we can put $\nu=0$ in the function $E^{-\nu}_{l}(r_{20})$
obtaining 
\beq
\int \frac{d^{2}R\exp(ilR)r_{20}}{(2\pi)^{2}|R+r_{20}/2||R-r_{20}/2|}\equiv
F(l)
\eeq
The rest of the Green function is easily calculated by
the stationary point method to finally give
\beq
\chi_{2}(s,l,q)=8\sqrt{\pi}s^{\Delta}(a\ln s)^{-3/2}F_{2}(l)J_{2}(l,q)
\eeq
where
\beq
\Delta= \omega_{0}=(g^{2}N/\pi^{2})\ln 2,\ \
a=(7g^{2}N/2\pi^{2})\zeta (3)
\eeq

\section{The projectile part}
The BFKL Green function at $l=0$, which enters (12)
is given by the expression [8]
\beq
G_{0}(s,r,r')=(1/8)rr'\int_{-\infty}^{\infty}\frac{d\nu s^{\omega(\nu)}}
{(\nu^{2}+1/4)^{2}}(r/r')^{-2i\nu}
\eeq
 Applying to (24) the operator $\nabla^{4}$ we obtain
\beq
\nabla^{4}G_{0}(s,r,r')=(2r'/r^{3})\int_{-\infty}^{\infty}
d\nu s^{\omega(\nu)}(r/r')^{-2i\nu}
\eeq

Before integrating over $r$ in (12) we shall integrate over
the energy $s_{1}$, as indicated in (11). Part of the $s_{1}$-dependence
comes from  factors $\chi_{2}$. Recall that $s_{2}=s/s_{1}$ and that also
$s>>s_{1}$. This latter condition allows to neglect the dependence on
$s_{1}$ in the logarithmic factor in (22), so that the two factors
$\chi_{2}$ will only contribute a factor $s_{1}^{-2\Delta}$. We then obtain
an integral
\beq
\int ds_{1}s_{1}^{-1-2\Delta}\nabla^{4}G_{0}(s,r,r')=
(2r'/r^{3})\int_{-\infty}^{\infty}
d\nu (r/r')^{-2i\nu}(2\Delta-\omega(\nu))^{-1}
\eeq

Now we have to finally integrate over $r$ to obtain the function $\chi_{1}$,
Eq. (12), integrated over $s_{1}$. This integration requires some care due
to a high singularity of the righthand side of (26) at $r=0$. To do it, we
consider $\Delta$ in (26) as a variable and first take $\Delta<0$. Then the
denominator in (26) will not vanish in the strip of the upper half-plane of
$\nu$ with ${\mbox Im}\,\nu<\frac{1}{2}+\epsilon$, $\epsilon>0$. This
allows to shift the integration contour in (26) to a line
${\mbox Im}\,\nu=\frac{1}{2}+\epsilon$. Then the singularity in $r$ will be
diminished to make the integration over $r$ possible. We get
\beq
J_{1}(q,r_{1})\equiv\int ds_{1}s_{1}^{-1-2\Delta}\chi_{1}(s_{1},q,r_{1})=
-\pi q r_{1}\int_{{\mbox Im}\,\nu=\frac{1}{2}+\epsilon}
d\nu (qr_{1}/2)^{2i\nu}\frac{1}{2\Delta-\omega(\nu))}\frac
{\Gamma(1/2-i\nu)}{\Gamma(1/2+i\nu)}
\eeq
With $\Delta<0$ the integrand on the righthand side has no singularities in
the strip ${\mbox Im}\,\nu<\frac{1}{2}+\epsilon$, since the pole of the
$\Gamma$ function at $\nu=i/2$ is compensated by the pole of the function
$\psi$ in $\omega(\nu)$ at the same point. So we can return to the
integration over the real $\nu$ and subsequently pass to the physical
value $\Delta>0$. Then finally 
\beq
J_{1}(q,r_{1})=
-\pi q r_{1}\int_{-\infty}^{+\infty}
d\nu (qr_{1}/2)^{2i\nu}\frac{1}{2\Delta-\omega(\nu))}\frac
{\Gamma(1/2-i\nu)}{\Gamma(1/2+i\nu)}
\eeq

This integral can be calculated as a sum of residues of the integrand at
points $\nu=\pm ix_{k}$, $0<x_{1}<x_{2}<...$, at which
\beq
2\Delta-\omega(\nu)=0
\eeq
Residues in the upper semiplane are to be taken if $qr_{1}/2>1$ and those
in the lower semiplane if $qr_{1}/2<1$. Thus we obtain
\beq
J_{1}(q,r_{1})=-\frac{(2\pi)^{4}}{g^{2}N}\sum_{k}a_{k}
(qr_{1}/2)^{1\pm 2x_{k}}\frac{\Gamma(1/2\mp x_{k})}{\Gamma(1/2\pm x_{k})}
\eeq
where 
\beq
a_{k}=
\frac{1}{\psi'(1/2-x_{k})-\psi'(1/2+x_{k})}
\eeq
and the signs should be chosen to always have $(qr_{1}/2)^{\pm 2x_{k}}<1$.

The first three roots of Eq. (29) are
\beq 
x_{1}=0.2648,\ \ x_{2}=1.3505,\ \ x_{3}=2.3704
\eeq
with the corresponding coefficients $a_{k}$
\beq
a_{1}=0.05944,\ \  
a_{2}=0.02139,\ \  
a_{3}=0.01610,\ \ 
\eeq

\section{The triple pomeron cross-section}
As we have found, the projectile part gives a contribution to the
cross-section (12) in the form of a sum of powers $(qr_{1})^{1\pm
2x_{k}}$, with rising values of $x_{k}$. For large $Q^{2}$ we expect that
$r_{1}\sim 1/Q$, so that the product $qr_{1}$ is small. Then the plus sign
should be taken in (30) and the bulk of the contribution is expected to
come from the lowest power $x_{1}$. For this reason in the following we
shall study the contribution from only the nearest pole $\nu=-ix_{1}$ to
(28), taking
\beq
J_{1}(q,r_{1})\simeq -\frac{(2\pi)^{4}}{g^{2}N}a_{1}
(qr_{1}/2)^{1+ 2x_{1}}\frac{\Gamma(1/2- x_{1})}{\Gamma(1/2+ x_{1})}
\eeq

Combining all the factors, in this approximation we find for the cross-section $\sigma$, Eq.
(11)
\beq
\sigma=\frac{g^{8}N^{2}}{(2\pi)^{6}}2^{-2x_{1}}a_{1}
B_{1}\frac{\Gamma(1/2- x_{1})}{\Gamma(1/2+ x_{1})}
\frac{s^{2\Delta}}{(a\ln s)^{3}}\langle r_{1}^{1+2x_{1}}\rangle
\int d^{2}l \,l^{-1+2x_{1}}F^{2}(l)
\eeq
Here we have introduced the average value of $r_{1}^{1+2x_{1}}$ in the
projectile 
\beq
\langle r_{1}^{1+2x_{1}}\rangle=\int d^{2}r_{1}
r_{1}^{1+2x_{1}}\rho_{1}(r_{1})
\eeq
$B_{1}$ is a number defined as
as a result of the $q$ integration
\beq
B_{1}=l^{1-2x_{1}}
\int (d^{2}q/(2\pi)^{2})|l/2+q|^{1+2x_{1}}(\nabla_{q}J(l,q))^{2}
\eeq
It does not depend on $l$ and can be represented as an integral over
three momenta
\beq
B_{1}=\frac{1}{(2\pi )^{4}l^{1+2x_{1}}}\int d^{2}qd^{2}pd^{2}p'
\frac{|l/2+q|^{1+2x_{1}}(q+p)(q+p')(l+p_{+}-p_{-})(l+p'_{+}-p'_{-})}
{p_{+}p_{-}p'_{+}p'_{-}|q+p|^{3}|q+p'|^{3}}
\eeq
where
 \beq p_{\pm}=|p\pm l/2|;\ \ p'_{\pm}=|p'\pm l/2|\eeq
It is a well-defined integral.  $B_{1}$
generalizes a similar constant $B$ which appears in the asymptotic triple
pomeron vertex [6], in which $x_{1}$ is absent. Calculations give
\[ B_{1}=6.84 \]

The explicit form of $F(l)$ is given by (21). So its square introduces
two more integrations, over $R$ and $R'$. Integration over $l$  then gives
\beq
\int d^{2}l\,l^{-1+2x_{1}}\exp il(R-R')=
 2^{1+2x_{1}}\pi\frac{\Gamma(1/2+ x_{1})}{\Gamma(1/2- x_{1})}
|R-R'|^{-1-2x_{1}}
\eeq 
We are left with the final integral over $R$ and $R'$
\beq
\int \frac{d^{2}Rd^{2}R'}
{(2\pi)^{4}|R-R'|^{1+2x_{1}}}\frac{1}
{|R+r_{20}/2||R-r_{20}/2||R'+r_{20}/2||R'-r_{20}/2|}=
\frac{D}{r_{20}^{1+2x_{1}}}
\eeq
which defines another numerical constant $D$. Numerical integration gives
\[ D=0.566 \]

Putting this into (35) we obtain the final cross-section
\beq
\sigma=\frac{g^{8}N^{2}}{(2\pi)^{5}}a_{1}B_{1}D
\frac{s^{2\Delta}}{(a\ln s)^{3}}r_{20}^{1-2x_{1}}\langle
r_{1}^{1+2x_{1}}\rangle
\eeq
with the known numerical constants $a_{1}$, $B_{1}$ and
$D$.

 The $Q^{2}$-dependence of the
cross-section (42) is concentrated in the average value 
$\langle r_{1}^{1+2x_{1}}\rangle$. At large $Q^{2}$ this average has the
order $Q^{-1-2x_{1}}$, which determines the order of the cross-section
$\sigma$ to be
\beq \sigma\sim (r_{20}/Q)(Qr_{20})^{-2x_{1}}\eeq
This should be compared with the cross-section which results from the 
pure two-pomeron exchange, Fig. 1$b$. As found in [1], it has the same
dependence on $s$ but falls only as $1/Q$ at large $Q^{2}$. Therefore for
large $Q^{2}$ the triple pomeron contribution is neglegible relative to the
pure two-pomeron exchange, due to the reduction of the anomalous dimension
by $2x_{1}$. This is the main result of our study.

Two comments are to be added in conclusion. First, one might think that the
obtained result is only a consequence of different scales of the projectile
and target. Taking the virtality of the target of the same order $Q^{2}$,
one might argue that $\sigma\sim 1/Q^{2}$ on dimensional grounds, which is
of the same order as for the pure two pomeron exchange. However this
argument would be wrong. With a highly virtual target, one cannot simply
put $r'_{2}=r'_{3}=r_{20}$, but has to consider the perturbative density
$\rho_{2}$, which is singular at the origin. Then one has to regularize the
integrations in $r'_{2}$ and $r'_{3}$ by introducing a finite  mass $m$
for quarks inside the target. As a result, the average values of
$r'_{2,3}$ will not have the order $1/Q$ but rather $1/m$. Then the final
conclusion will remain the same: the contribution will be damped by the
factor $(m/Q)^{2x_{1}}$.

Second, the procedure followed here for the nearest pole at $\nu=-ix_{1}$
cannot be trivially generalized to other poles. The point is that with
$x_{k}>1/2$ neither the integral (38) nor the integral (41) converge. For
such $x_{k}$ one has to use the general form (30), taking different signs in
different parts of the ($q, r_{1}$) phase space. Then the integrals over
$r_{1},\,q,\,R$ and $R'$ do not decouple and the cross-section turns out to
be represented by a very complicated 12-dimensional integral over
$r_{1},\,q,\,R,\,R',\,p$ and $p'$. However one can estimate the resulting
$Q^{2}$ dependence by noting that for $x_{k}>1/2$ the extra dimension in
(38) will be supplied by the corresponding power of $Q$. One then finds that
for all $x_{k}>1/2$ the cross-section has the same order $1/Q^{2}$. Thus,
although the contributions of all $x_{k}>1/2$ are definitely smaller than
the one from $x_{1}$, they all have to be calculated simultaneously.
Therefore our derivation is only practical for the dominant contribution
corresponding to the pole at $\nu=-ix_{1}$.

\section{Conclusions.}
We have shown that for highly virtual probes the triple (and hopefully
multiple) pomeron interaction does not contribute to the cross-section, so
that it can be calculated as a sum of independent multipomeron exchanges,
as has been done in [1]. The amplitude then aquires an essentially eikonal
form.

For colliding low-mass particles the triple pomeron does contribute
significantly. We have not been able to find this contribution in a form
suitable for practical calculations. However we would like to stress that
even if we had succeded, that would not have solved the problem. On the one
hand, when
a higher number of pomerons is exchanged
higher order multipomeron interactions evidently come into play, whose
calculation is still more hopeless. On the other hand, for low-mass
particles the coupling to a pomeron (or to several pomerons) is
essentially non-perturbative. So even without any multipomeron interactions
calculation of multipomeron exchanges becomes hardly possible.
Multipomeron interactions then  do not make the situation significantly
worse, only adding a contribution which can be calculated perurbatively, in
principle, but not in practice. 

\section{Acknowledgements}
The author expresses his deep gratitude to Prof. Carlos Pajares for his
constant interest in the present work and helpful discussions. He is
especially thankful to Dr. Gavin Salam, whose comments have initiated this
study. He also thanks IBERDROLA, Spain,  for financial support.

\newpage
\section{References}
\noi 1. M.A.Braun, Univ. of St. Petersburg preprint SPbU-IP-1995/3
(hep-ph/9502403, to be published in Z. Phys. {\bf C}).\\
2.  V.S.Fadin, E.A.Kuraev and L.N.Lipatov, Phys. Lett. {\bf B60} (1975)
50; I.I.Balitsky and L.N.Lipatov, Sov. J. Nucl. Phys. {\bf 15} (1978) 438.\\
3. A.H. Mueller, Nucl. Phys. {\bf B415} (1994) 373.\\
4. A.H.Mueller and B.Patel, Nucl. Phys., {\bf B425} (1994) 471.\\
5. A.H. Mueller, Nucl. Phys. {\bf B437} (1995) 107.\\
6. M.A.Braun, Univ. of St. Petersburg preprint SPbU-IP-1995/10
(hep-ph/9506245, to be published in Z.Physik {\bf C}).\\
7. J.Bartels, Nucl. Phys. {\bf B175} (1980) 365.\\
8. L.N.Lipatov, in {\it Perturbative Quantum Chromodynamics},
Ed. A.H.Mueller, Advanced Series
on Directions in High Energy Physics, World Scientific, Singapore 1989.\\
9. A.H.Mueller and W.K.-Tang, Phys. Lett. {\bf B 284} (1992) 123.\\

\newpage
\section{Figure captions}
\noi 1. Triple pomeron ($a$) and "pure" double pomeron exchange ($b$)
contributions to the scattering amplitude.\\
2. The generic double pomeron exchange diagram for the scattering amplitude.

%%%%%%%%%%%%%%%%%%%%%%%%%%%%%%%%%%%%%%%%%%%%%%%%%%%%%%%%%%%%%%%%%

%Figure 1.
\newpage
\begin{picture}(300,300)(0,-50)
\thicklines
\put (25,225){\line (1,0){100}}
\put (25,25){\line (1,0){100}}
\put (73,225){\line (0,-1){100}}
\put (77,225){\line (0,-1){100}}
\put (75,75){\oval(50,100)}
\put (75,75){\oval(42,92)}
\put(75,225){\circle*{8}}
\put(75,125){\circle*{8}}
\put(75,25){\circle*{8}}
\put(75,0){\makebox(0,0){\Large a}}
\put(150,-20){\makebox(0,0){\Large Fig. 1}}

\put (175,225){\line (1,0){100}}
\put (175,25){\line (1,0){100}}
\put (225,125){\oval(50,200)}
\put (225,125){\oval(42,192)}
\put(225,225){\circle*{8}}
\put(225,25){\circle*{8}}
\put(225,0){\makebox(0,0){\Large b}}
\end{picture}

%Figure 2.

\newpage
\begin{picture}(150,300)(0,-50)
\thicklines
\put (15,225){\line (1,0){120}}
\put (25,25){\line (1,0){100}}
\put (75,125){\oval(50,200)[b]}
\put (75,125){\oval(42,192)[b]}
\put(35,125){\framebox(80,100){\Large\bf B}}
\put(75,25){\circle*{8}}
\put(75,-20){\makebox(0,0){\Large Fig. 2}}	
\end{picture}

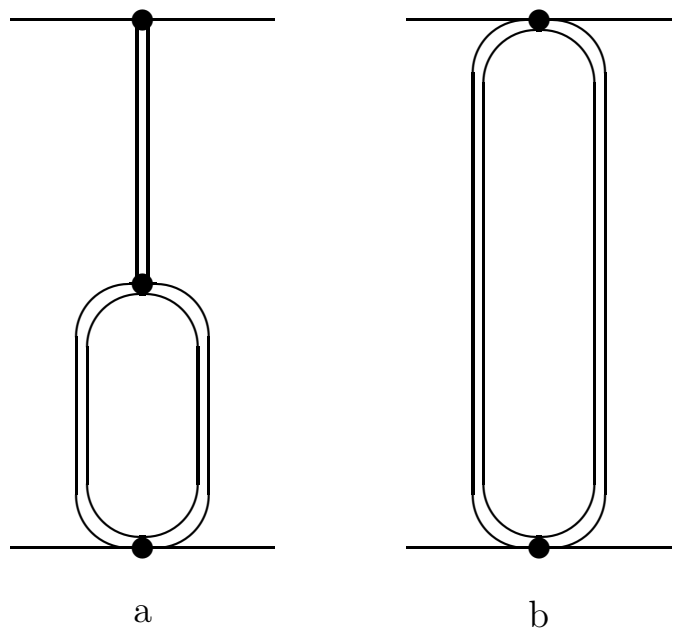
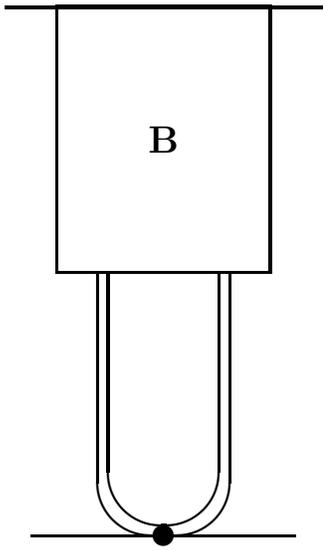
\end{document}